\documentclass[12pt]{article}
\usepackage{graphicx}
\newcommand{\beq}{\begin{equation}}
\newcommand{\eeq}{\end{equation}}
\newcommand{\bea}{\begin{eqnarray}}
\newcommand{\eea}{\end{eqnarray}}

\newcommand{\be}{\begin{equation}}
\newcommand{\ee}{\end{equation}}
\newcommand{\ba}{\begin{eqnarray}}
\newcommand{\ea}{\end{eqnarray}}

\begin{document}
\begin{titlepage}
\pagestyle{empty}
\vspace{1.0in}
\begin{flushright}
\end{flushright}
\vspace{.1in}
\begin{center}
\begin{large}
{\bf Duality in $SU(N) \times SU(N')$ product group from M theory}
\end{large}
\vskip 0.5in
Julian Lee$^{+}$ \\
\vskip 0.2in
{\small{\it Korea Institute for Advanced Study\\  Seoul 130-012, Korea}}\\
and \\
\vskip 0.2in
Sang-Jin Sin$^{*}$ \\
\vskip 0.2in
{\small {\it Department of Physics, Hanyang University \\  Seoul 133-791,
Korea}}
\end{center}

\begin{abstract}
We generalize the M-theoretic duality of Schmaltz and Sundrum to the
product group $SU(N)\times SU(N')$ case.  We show that the
type IIA brane configurations for dual gauge theories are in
fact two special limits of the same M-theory 5-brane,
just as in the case of the simple $SU(N)$ group.
\end{abstract}

\vspace{.3in}

\vspace{5cm}
\noindent $^{+}$: jul@kias.re.kr \\
\noindent $^{*}$: sjs@dirac.hanyang.ac.kr
\end{titlepage}

\section{ Introduction}
Recently there have been many attempts to understand  properties of
supersymmetric gauge theories using branes in string theory and
M-theory\cite{gk}-\cite{epr}.
One of the properties which are difficult to understand from the field
theory point of view but can be easily understood from the geometric
configurations in the brane picture  is
electric-magnetic duality\cite{sei}-\cite{ils}. It was first demonstrated that
along with other properties, the duality in three dimensional $N=1$ $SU(N)$
super Yang-Mills (SYM) theory can be derived using weakly coupled type IIB
string brane picture\cite{hw}.  Starting from the string brane setup
corresponding to a given
gauge theory, the configuration for the dual gauge group could be obtained
through the movement of branes. This consideration was subsequently extended to
the SYM theory in four dimensions\cite{ba}-\cite{bsty}, using the type IIA
string branes. In particular, the duality
for the case of product gauge group such as $U(N) \times U(N')$ or $SO(N) \times
SO(N')$ in the type IIA string picture was discussed in
papers such as ref.\cite{bh}-\cite{ta}.

On the other hand, supersymmetric four dimensional gauge theories can also be
investigated using M-theory branes \cite{witten}-\cite{epr}. The
brane setup one considers is a single smooth M 5-brane, which
becomes type IIA string branes in the weak coupling limit.
 It has an
advantage over string branes in that it contains quantum informations. After
Witten discussed various properties of $N=2$ and $N=1$ SYM theory using M-theory
5 brane
picture\cite{witten}, similar investigations were done for other gauge
theories\cite{hoo}-\cite{epr}, and the curves
for general product group were written in \cite{nos,gp}.
The discussion of duality using M5 brane was first done by Schmaltz and
Sundrum\cite{ss}\footnote{see also ref.\cite{sug}} for the case of gauge group
$SU(N)$. The dual configurations appear as smooth deformations of one another
in contrast to weakly coupled  string theory where singular situations
arise at the intermediate stages of the deformations.  This was subsequently
generalized to other simple groups such as  $Sp(N)$ and $SO(N)$ \cite{ca}.

The purpose of this paper is to extend these ideas of M-theoretic duality
 to the case of theories with product gauge group.
  We show that the
type IIA brane configurations for dual gauge theories are in
fact two special limits of the same M-theory 5-brane,
just as in the simple $SU(N)$ case. 

The organization of this paper is as follows. Section 2 reviews the
MQCD duality for the $SU(N)$ case. In section 3, we extend
 the MQCD  duality  to the product group $SU(N) \times
SU(N')$, which is the main result of our paper. Section 4 contains concluding remarks.

\section{Duality in $SU(N)$ theories}
\subsection{The duality in Gauge theory}
The dual of the SU(N) gauge theory   is known to be SU(F-N) theory
with F being the number of the flavors. The non-perturbative
superpotential is known to be\cite{ads}
\be
W_{{\rm {eff}}}= (N - F) \left(\frac{\Lambda _{N=1}^{3N - F}}{\det M
}\right)^{1/(N -F)}   + {\rm Tr}(m_f M) \label{gaugepot},
\ee
where $M=Q\tilde{Q}$ is the meson field and $m_f$ is the mass matrix.
We assume that all $m_i$ are non-zero throughout this paper.
Now,  one interesting aspect is that the minimum of the superpotential
is obtained when
\be
m_f M=  \Lambda^{3-F/N}(\det m_f)^{1/N}:=\zeta.  \label{minimum}
\ee
In the dual theory, the tree level superpotential is given by
 \be
 W_{tree}=\frac{1}{\mu} M_{ij}q_i{\bar q}_j + {\rm Tr}(m_f M)
\ee
where $M$ is now an independent singlet field.
We see that the vacuum expectation value of the meson variable $M$
plays the role of the dual quark mass
\be
{\tilde m}=\frac{M}{\mu}=\frac{\zeta}{\mu}(m_f)^{-1}. \label{dualmass}
\ee

\subsection{The duality in string brane picture}
 First consider the brane configuration in the weakly coupled type IIA
string theory\cite{ss}. Let us denote the ten dimensional spacetime coordinates as
$x^0, x^1, \cdots x^9$, where $x^0,\cdots x^3$ denote the usual 4-d
spacetime where the gauge theory lives. There are two NS branes, with $N$
D4 branes stretching between them, and $F$ semi-infinite D4 branes to one
of the NS branes. The 4-d spacetime is shared by all the branes, and the
remaining 6 dimensional space becomes the internal space from the field
theory point of view. It is convenient to introduce the complex
coordinates $v=x^4 + i x^5$ and $w= x^8 + i x^9$.  The left NS brane,
which we denote A, spans the $v$ plane while the other one, which we call B,
spans $w$ plane.  The magnetic dual of this configuration is given by
 brane movements and reconnections \cite{hw} - \cite{ta}.
The equivalence of the dual pair is the based on the assumption that the
infrared physics does not change during this process.

\subsection{The duality in M-theory}
In M-theory the web of branes described above is replaced by a
single M5 brane\cite{witten, hoo}.
The M-brane  in our context is described by the curve
\begin{eqnarray}
w= &&{\frac{\zeta}{v }}  \nonumber \\
t= &&\frac{\xi v^{N}}{\prod_{i=1}^{F}(v-m_{i})} \label{curve}
\end{eqnarray}
where  $t=\exp(-(x^6+ix^{10})/R)$ with $R=g_sl_s$ being the radius of the
eleventh dimension, and $\zeta$ and $\xi$ are the parameters characterizing the curve. They can be related to the field theory parameters\cite{ss}:
\bea
\zeta &=&  \Lambda^{3-F/N}(\det m_f)^{1/N} \\
\xi &=& \prod(m_i)^{1-N/F}
\eea
The algebraic form of the curve is uniquely determined by the bending power and holomorphy. That is, we require $t \sim v^N $ for $ v \to 0$ and $t \sim v^{F-N}$ for $v \to \infty$. Then the form of the holomorphic curve with this condition is uniquely given by (\ref{curve}). The bending power can be obtained by considering {\it either} the electric or magnetic brane configuration (but not necessarily both). Then the corresponding M-theory curve (\ref{curve}) automatically contains the dual description, as we will review below.  

Now consider the perturbative string theory limit, $R\to 0$. From here on, we take all the quantities dimensionless by dividing by the string scale $l_s$.
As we take this limit, we keep the distances between two would-be NS5 branes fixed. Then depending on whether we keep the electric or magnetic quark masses finite, we get the electric or magnetic configuration respectively. To be more quantitative, since each of the would-be NS5 branes spans the $v$ plane and $w$ plane respectively, we define their positions by
\bea
s_A &\equiv&  -R\log t(v=v_A)
\nonumber \\
s_B &\equiv&  -R \log t(w=w_B)
\eea
where $v_A$ and $w_B$ are some fixed numbers. The final results does not depend on the arbitrary choice of  $v_A,w_B$. The idea is that as we take
the string theory limit, the dominant part becomes\cite{ss}
\beq
\Delta s \equiv s_A - s_B \simeq -RN \log \zeta
\eeq
if we keep the electric quark masses $m_i$'s fixed.  Since we want to fix
the the separation of the branes $| \Delta s |$ to be finite, we see that $\zeta\to 0 $.
We also note that in this limit,
\bea
\Delta x_6 &=& \Re(s) \simeq -RN \log |\zeta | > 0 , \\
\Delta x_{10} &=& \Im(s) \simeq -RN \arg (\zeta) \to 0 ,
\eea
that is, the A brane goes to the left of the B brane and the eleventh
dimension vanishes.
The magnetic configuration is obtained if we take the limit $R \to 0$ while keeping  dual magnetic quark mass $\tilde m_i \equiv {\zeta \over m_i}$'s finite.  We then have
\beq
\Delta s \simeq R(F-N) \log \zeta
\eeq
in this limit. Notice that  here  $\Delta x_6 < 0$ for $F>N$,
which shows that the positions of A and B brane get interchanged for
$F>N$ 
  
One can also check that the number of D4 branes between two NS5 branes come out correctly: when $t$ takes the value between
the two (to be ) NS5 branes, for example if we take $s={s_A + s_B \over 2}$, then corresponding $v$ is of order $O(\zeta^{1/2})$. For the electric configuration where $m_i$'s are kept finite, the $v-t$ relation in this region becomes
\be
t\sim v^N,
\ee
 which implies that there are N D-branes for fixed  $t$. Hence the gauge
 group is $SU(N)$.
 On the other hand, in the magnetic configuration where the values of $m_i$'s taken to be order of  $\zeta$, the $t-v$ relation for degenerate region becomes
\be
t\sim v^{-(F-N)},\ee
 using $v\sim O(\sqrt{\zeta})>>\zeta $.
Therefore the curve represent the M-theory configuration for gauge group $SU(F-N)$. This can be seen more readily by
rewriting the curve\cite{ss} in the meson picture\cite{hoo}:
\begin{eqnarray}
v &=&\frac{\zeta }{w}  \nonumber \\
t &=&\frac{\xi_D w^{F-N}}{\prod_{i=1}^F (w- \tilde m_i)} \label{dualc}
\end{eqnarray}
where $\xi_D =\xi \zeta^{N-F}\prod_i (-m_i)$,
and taking the perturbative limit $R \to 0$ with $\tilde m_i \equiv {\zeta \over m_i}$ fixed. Note that even before taking the perturbative string limit, one can easily realize that the same M-theory curve which corresponds to the electric theory with $SU(N)$ gauge group in a certain limit can also describe the magnetic configuration with $SU(F-N)$ gauge group in another limit, since (\ref{curve}) and (\ref{dualc}), which is nothing but the same curve written in different variables, are exactly of the same form, with $N$ replaced by $F-N$. This is closely related to the fact that the string brane configurations for these two cases have exactly the same form,with $v$ and $w$, $N$ and $F-N$ exchanged with each other. However, the string brane configuration for electric and magnetic cases have totally different forms for the product group model we will be considering\cite{bh}. Therefore we cannot expect that the dual gauge group can be easily read off simply by changing variables. In this case it is important to take the limit carefully and explicitly count the number of intermediate D4 branes, as was done in this section for simple $SU(N)$.

Summarizing,
when we take the string theory limit, the curve (\ref{curve}) for  finite $m_i$'s,
is reduced to the electric D-brane configuration,
while   for $m_i\sim O(\zeta)$, or equivalently for finite $\tilde m_i$,
the same curve is reduced to the magnetic brane configuration. 

\section{Duality in $SU(N)\times SU(N')$ model}
\subsection{Duality in Gauge Theory}
In this section, we consider a model with product group. The electric
configuration is given by $SU(N) \times SU(N')$ with $F(F')$ flavors of quarks
and anti-quarks   $Q,\tilde Q (Q', \tilde Q')$ in the fundamental and its
conjugate representation of $SU(N) (SU(N'))$, the adjoint fields $A$, $A'$, 
 a bifundamental and its conjugate $X, \tilde X$.  We consider the tree level superpotential given by
\beq
W_{tree}={\rm Tr}{X A \tilde X} + {\rm Tr}{\tilde X A'  X}+ {m_1 \over 2} {\rm Tr} A^2 + {m_2 \over 2}{\rm Tr} A'^2  \label{supo}
\eeq
which reduces to 
\beq
W_{tree}=({1 \over m_1} + {1 \over m_2}){\rm Tr}(X \tilde X)^2. \label{supone}
\eeq
after integrating out $A$,$A'$. Then the coefficient can be set to one by rescaling $X$,$\tilde X$, when it is nonzero.
  As was discussed in ref.\cite{ils}, the magnetic dual for this theory has
gauge group $SU(\tilde N) \times SU(\tilde N')$ where $\tilde N = 2F'+F-N'$ and
$\tilde N' = 2F+F'-N$, with $F'(F)$ flavors of magnetic quarks and anti-quarks
$q,\tilde q (q', \tilde q')$ in the fundamental and its conjugate representation
of $SU(\tilde N) (SU(\tilde N'))$, a bifundamental and its conjugate $Y, \tilde
Y$, the superpotential being
\bea
W_{tree}=&&{\rm Tr}(Y \tilde Y)^2+M_1 q'\tilde q' + M_0 q' \tilde Y Y \tilde
q'+M'_1 \tilde q q + M'_0 \tilde q Y \tilde Y q +P_1 q \tilde Y q'   \nonumber \\
&+& \tilde P_1 \tilde q' Y \tilde q . \label{suma}
\eea
where $M_0$, $P_0$, etc, are singlet meson fields. In deriving the duality using M-theory, we will consider a more general theory  where the superpotential (\ref{supo}) is deformed with additional terms\cite{bh,gp} 
\beq
W_{add}= {\rm Tr}(m \tilde Q Q) + {\rm Tr}(m' \tilde Q' Q') + \mu {\rm Tr} X \tilde X  + {\rm Tr}(\lambda Q A \tilde Q) + {\rm Tr}(\lambda' Q' A' \tilde Q') .
\eeq
Then we expect to have these terms in the magnetic superpotential (\ref{suma}) expressed in terms of appropriate singlet fields.\footnote{In the  weakly coupled string brane setup, one can derive duality only for the case where these deformation parameters are zero\cite{bh}, whereas in the case of M theory it is easier to consider the case when they are all nonzero. This is the feature also found in simple group cases\cite{ss,ca}. The undeformed case (all quarks massless) for simple $SU(N)$ was considered in ref.\cite{sug}.}

\begin{figure}
\begin{center}
\includegraphics[angle=0, width=80mm]{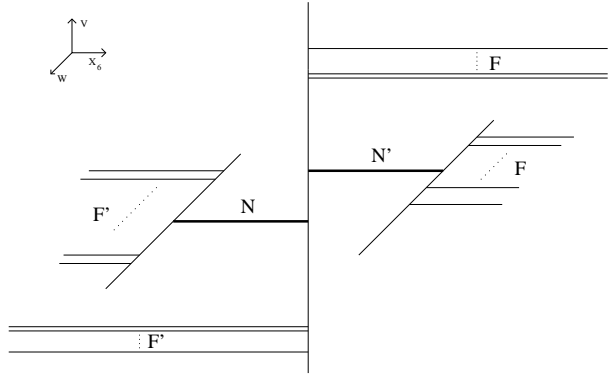}
\end{center}
\label{elep}
\caption{\it The classical string theory configuration corresponding to the
 electric theory, for $SU(N)\times SU(N')$ model.}
\vskip .2in
\end{figure}

\subsection{Dual Configurations in M-theory}
The IIA string brane configuration for electric case is as in Fig.1 ,
and the
magnetic configuration is in Fig.3.\cite{bh}
  We will label the NS branes by A, B, C, which are from left to right in the
electric picture. As in the case of simple $SU(N)$, the angles between neighboring NS branes gives mass to the adjoint fields\cite{bh,gp},and the positions of the semi-infinite and finite D4 branes parametrize the deformation parameters. In particular, the distance between finite D4 branes gives mass $\mu$ to the bifundamental field, and the distances of the semi-infinite branes  from the finite D4 branes are proportional quark masses $m_i$, $m_i'$. The remaining geometric parameters are related to $\lambda,\lambda'$.\footnote{In the original setup with D6 brane, $\lambda,\lambda'$ are related to the angle between the D6 branes and its neighboring NS branes\cite{bh}. In particular, when they are nonvanishing, one can also move D6 branes in the directions opposite to those given in ref.\cite{bh} to get the configuration given in \cite{gp}. However, we are interested in the deformations of the dual brane configurations given in ref.\cite{bh}. } For simplicity, we will assume that the B brane spans the
$v$ plane, whereas A, C branes lie in the $w$ plane.\footnote{ If one starts
from the configuration with D6 branes, then it is crucial  that A and C branes
are not parallel in order to derive duality. Also, the coefficient for the superpotential (\ref{supone}) vanishes when A and C branes are parallel. However, in the brane setup with semi-infinite D4 branes, there is not much qualitative difference whether we take them parallel or not,  so we took this configuration to simplify the form of the curve. Maybe one can consider this as a sort of limiting configuration where the angle between A and C brane is very small. Of course, the essence of the subsequent discussions does not change even if we take A and C branes not to be parallel, with some changes in the asymptotic boundary conditions and the algebraic form of the curve.}
We will take the $v$ coordinates of A, C to be $0$,$1$ for convenience, other
choices corresponding to simple rescalings of $v$.
Then the asymptotic behavior of  the
corresponding M-brane is given by

 1) 2F' semi-infinite brane to the left

 2) 2F semi-infinite brane to right.

 3) NS branes have the following bending behavior:
 \bea
A: v \to 0 , \qquad w \to \infty , \qquad &&t \simeq { w^{N-F'} \over c}  \\
B: v \to \infty , \qquad w \to 0 , \qquad &&t \simeq v^{N'-N+F-F'}  \\
C: v \to 1 , \qquad w \to \infty , \qquad &&t \simeq  c' w^{F-N'}
\eea

As in the simple $SU(N)$ case, the asymptotic bending power is enough to
determine the curve along with the holomorphy. Also, depending on where we
attach
semi-infinite branes in the weakly coupled string limit, the gauge group given
by the
intermediate D4 branes is determined.
 However we note that  only $v$ can be used as a global
coordinate in contrast to the $SU(N)$ case.  The algebraic form of the
curve\cite{nos} with the asymptotic behavior above is given by
\bea
t &=& { v^{F'-N}(v
-1)^{N'-F}\prod_{i=1}^{2F}(v-v_i)  \over  \prod_{k=1}^{2F'}(v-v'_k)}  \label{ma}
\\
w &=& {\zeta \over v} + {\zeta' \over v-1}  \label{mb}
\eea
where  the zeroes and poles $v_i$ and $v'_l$ are the $v$ coordinate values of the
semi-infinite D4 branes.

We can now express the coefficients $c$, $c'$ appearing in the $t-w$ or $t-v$
relation at the asymptotic infinity in terms of the parameters  in  the
equations (\ref{ma}),(\ref{mb}):
\bea
c &=& {\prod_{k=1}^{2F'} (-v'_k) \over \prod_{i=1}^{2F} (-v_i)  } (-1)^{F-N'}
\zeta^{N-F'} \nonumber  \\
c' &=&  { \prod_{i=1}^{2F} (1-v_i) \over \prod_{k=1}^{2F'} (1-v'_k)}
\zeta'^{N'-F} \label{dth}
\eea

 The distances between NS branes are defined in exactly the same way as in the
simple $SU(N)$ case. Already, from the asymptotic behavior 1),2),3) and the
definition $s=-R \log t$, we easily see that in the weakly coupled limit $R \to
0$, the positions of A,C branes in $s$ direction are given by
\bea
s_A &\simeq& R \log c \label{do} \\
s_C &\simeq& -R \log c' \label{dt}
\eea
with $c$, $c'$ given by the equation (\ref{dth}).
Since the position of the middle NS brane labeled B is located at $s_B=0$ by
construction, we see that (\ref{do}),(\ref{dt}) give the distances between NS
branes.

\begin{figure}
\begin{center}
\includegraphics[angle=0, width=70mm]{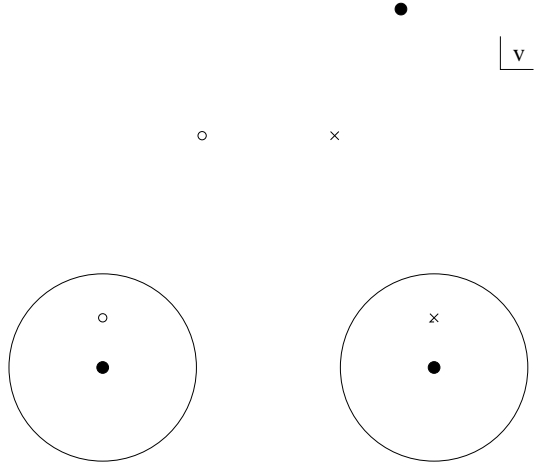}
\end{center}
\label{elepole}
\caption{\it Positions of poles corresponding to the semi-infinite D4 branes in
the
$v$ plane, in the electric limit. The black dots represent the poles
corresponding to the NS branes, which are located at $v=0, 1, \infty$. A cross
represents $F$ zero's of $t$, and a circle represents $F'$ poles of $t$. The
radii of the big circles are of order $\zeta.$}
\vskip .3in
\end{figure}

  The discussion of the weakly coupled limit is similar to the case of simple
$SU(N)$
case. This limit is defined as the one where $R$ goes to $0$,  and in order to
keep the
distances between A,B,C branes nonzero and finite,   we must send $\zeta$,
$\zeta'$ to $0$.\footnote{Again, $\zeta \to \infty$ or $\zeta' \to \infty$ are
not sensible limits to take, just as $\zeta \to \infty$ limit in the simple
$SU(N)$ case, since $v-w$ curve blows up.}  We will assume they are of the same
order as they approach zero, and this will be denoted by $O(\zeta)$. Just as in
the case of simple SU(N), we have to decide where we
attach the semi-infinite D4 branes in this limit. Depending on where and how
fast we send the semi-infinite D4 branes, the relative order of NS branes in
$x_6$ direction and the number of degenerate finite  D4 branes between them are
determined, as will be shown next.

\subsubsection{electric configuration}
In the weakly coupled electric limit, we have $F'$ left semi-infinite D4 branes
attached to A brane, $F'$ of them on B, $F$ of right semi-infinite D4 branes on
B and $F$ of them on C brane(Fig.2). That is:
\bea
A: v'_k &\sim& O(\zeta) ,     \qquad (k=1 \cdots F') \nonumber \\
B: v'_k &\to& {\rm finite},   \qquad (k= F'+1 \cdots 2F') \nonumber \\
B: v_i &=& {\rm finite},  (i= 1 \cdots F) \nonumber \\
C: v_i &\sim& 1+O(\zeta),   \qquad (i= F+1 \cdots 2F)
\label{pol}
\eea
where we also used the conditions that the corresponding $w$ values should stay
finite and nonzero as $\zeta \sim \zeta' \to 0$.

   Substituting these results into Eq.(\ref{do}), (\ref{dt}), (\ref{dth}), we
get
\bea
s_A &\simeq& NR \log \zeta \label{da} \\
s_C &\simeq& -N'R \log \zeta \label{dc}
\eea
and since $\log \zeta < 0$,  we have A,B,C NS branes from left to right
(Fig.1).

 The gauge group can be read off by counting the number of degenerate finite D4
branes stretched between the NS branes. The semi-infinite D4 branes  attached to
the middle B brane are separated in $v$ space, so they will not be included.
As was done in $SU(N)$ case, take any point A' between A and B branes. For example, if we take
\beq
s_{A'} = {s_A \over 2},
\eeq
then the
corresponding $t$ coordinate  goes to infinity in the limit $R \to 0$ like $O(\zeta^{-N/2})$.\footnote{If we take $s_{A'} = {s_A \over m}$ with different $m>1$, then the corresponding $t$ coordinate is $O(\zeta^{-N/m})$ and the following arguments are similar. }  We then look for
the  {\it degenerate} solution of the equation
\beq
 t \sim \zeta^{-N/2} =  { v^{F'-N}(v
-1)^{N'-F}(v-1+O(\zeta))^F \prod_{i=1}^F (v-v_i)   \over  (v-O(\zeta))^{F'}
\prod_{k=F'+1}^{2F'}(v-v'_k) }\sim ({1 \over   v})^N  \label {deeq} .
\eeq
where the last relation holds near $v \sim \zeta^{1/2}$ where the degeneracy occurs.
These solutions  obviously represent $N$ degenerate finite D4 branes suspended
between A and B brane, which gives the gauge group $SU(N)$.
Using the exactly the same kind of argument one can easily show that the number
of finite D4 branes between B and C branes is $N'$.

Thus, we obtain the IIA string brane configuration corresponding to the electric
gauge group, as expected.

\begin{figure}
\begin{center}
\includegraphics[angle=0, width=80mm]{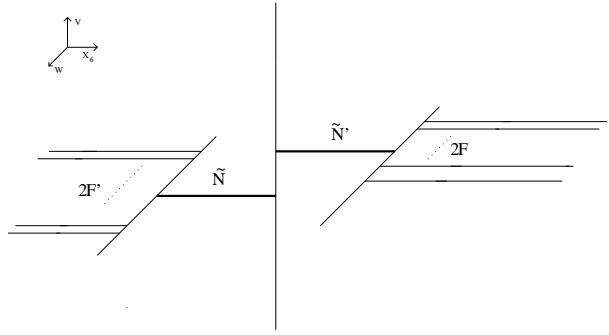}
\end{center}
\label{magp}
\caption{\it The classical string  theory configuration corresponding to the
magnetic theory, for $SU(N)\times SU(N')$ model.}
\vskip .2in
\end{figure}

\subsubsection{magnetic configuration}
 In this case we take
\bea
A:  v_i &\sim& O(\zeta) ,     \qquad (k=1 \cdots 2F) \nonumber \\
C: v'_k &\sim& 1+O(\zeta).   \qquad (i= 1 \cdots 2F')
\label{pool}
\eea
which is depicted in Fig.4.
  Substituting these results into Eq.  (\ref{dth}), (\ref{do}), (\ref{dt}), we
get
\bea
s_A &\simeq& -\tilde N'R \log \zeta \label{daa} \\
s_C &\simeq& \tilde N R \log \zeta \label{dcc}
\eea
where $\tilde N \equiv 2F'+F-N', \tilde N' \equiv 2F+F'-N$.
 Therefore, for $\tilde N, \tilde N' > 0$, which of course is the condition for
the magnetic theory to exist, the signs of $s_A$, $s_c$ are opposite to those in
the electric limit,  so there are C, B, A
NS branes from left to right.(Fig.3)  Substituting (\ref{pool}) into
(\ref{ma}), we now get for $t \sim \zeta^{\tilde N' /2}$, 
\beq
t  = { v^{F'-N}(v
-1)^{N'-F}\prod_{i=1}^{2F}(v-O(\zeta))  \over  \prod_{k=1}^{2F'}(v-1+O(\zeta) )}
\sim v^{\tilde N'}.
\eeq
where again, the last relation holds near $v \sim \zeta^{1/2}$, where the degeneracy
occurs. Thus the number of degenerate finite D4 branes is $\tilde N'$ and we
immediately  read off the gauge group to be $SU(\tilde N')$.     Using the
exactly same kind of argument, one can show that the number of degenerate D4
branes between C and B brane is $\tilde N$.

\begin{figure}
\begin{center}
\includegraphics[angle=0, width=70mm]{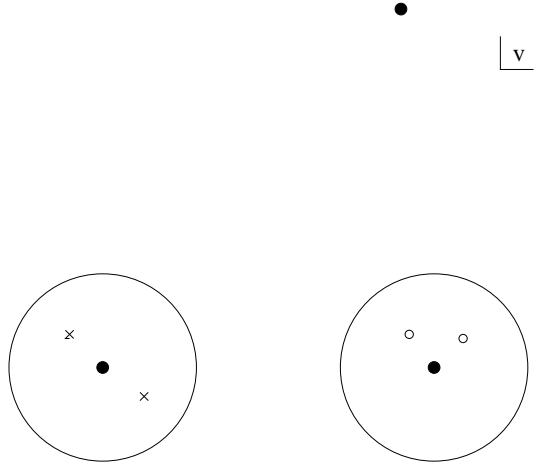}
\end{center}
\label{magpole}
\caption{\it Positions of poles corresponding to the semi-infinite D4 branes in
the
$v$ plane, in the magnetic limit.}
\vskip .2in
\end{figure}

\section{concluding remarks}
In this paper, we considered the duality for the
product group $SU(N) \times SU(N')$ in the M-theory 5 brane settings. 
We find that the dual configurations can be obtained as different limits of the same
M-theory branes. This is a generalization of the work done for the simple group cases.\cite{ss,sug,ca} However, we could not make a mapping between the parameters of
the M-theory curve and the those of the field theory, the main reason being that
not much is known about the quantum properties of the model we are considering.
It seems more investigations are needed in the field theory side in order to
have clearer understandings of these issues.

\vskip 1cm
\noindent{\bf \Large Acknowledgement}

\noindent We would like to thank  Chanyong Park for useful discussions.
The work of SJS has been supported by the research grant
KOSEF 971-0201-001-2.

\end{document}